\documentclass[sigconf]{acmart}

\AtBeginDocument{%
  }

\begin{document}


\title[Augmented Co-Speech Gesture Generation]{Augmented Co-Speech Gesture Generation: Including Form and Meaning Features to Guide Learning-Based Gesture Synthesis}


\author{Hendric Voß}
\email{hvoss@techfak.uni-bielefeld.de}
\affiliation{%
  \institution{Social Cognitive Systems Group}
  \institution{Bielefeld University}
  \streetaddress{Universitätsstraße 25}
  \country{Germany}
}

\author{Stefan Kopp}
\email{skopp@techfak.uni-bielefeld.de}
\affiliation{%
  \institution{Social Cognitive Systems Group}
  \institution{Bielefeld University}
  \streetaddress{Universitätsstraße 25}
  \country{Germany}
}

\renewcommand{\shortauthors}{Voß and Kopp}

\begin{abstract}
Due to their significance in human communication, the automatic generation of co-speech gestures in artificial embodied agents has received a lot of attention. Although modern deep learning approaches can generate realistic-looking conversational gestures from spoken language, they often lack the ability to convey meaningful information and generate contextually appropriate gestures. This paper presents an augmented approach to the generation of co-speech gestures that additionally takes into account given form and meaning features for the gestures.
Our framework effectively acquires this information from a small corpus with rich semantic annotations and a larger corpus without such information. We provide an analysis of the effects of distinctive feature targets and we report on a human rater evaluation study demonstrating that our framework achieves semantic coherence and person perception on the same level as human ground truth behavior. We make our data pipeline and the generation framework publicly available. 
\end{abstract}




\maketitle

\section{Introduction}
Human communication encompasses a wide range of expressive modalities, from verbal expressions and facial cues to bodily gestures. This spectrum of communicative signals enables us to convey complex messages and immerse ourselves in rich and meaningful interactions. Consequently, much work has been directed to machines that can process verbal and nonverbal signals to facilitate seamless human-technology interaction. One prominent challenge is the automatic generation of speech-accompanying hand gestures for embodied agents. 
Recent approaches apply machine learning to train models, that can predict gestures for a given linguistic input. This speech-driven synthesis has shown impressive results in generating realistic gestural motion aligned to a given spoken utterance or text. However, these techniques struggle with generating \textit{representational gestures} to convey meaningful information, as the required information for these gestures is not solely conveyed through speech. Further, it is hard to guide such trained models to generate specifically desired, contextually appropriate gestures.

We present a learning-based approach to imbue a state-of-the-art framework for co-speech gesture synthesis with the ability to generate gestures that are not only realistic but also semantically meaningful and contextually appropriate. To that end, we incorporate form and meaning target features (e.g. gesture category, handedness, or what entity it shall depict) into the gesture synthesis process. We demonstrate that our framework effectively acquires such form and meaning information from a small corpus with corresponding annotations and a substantially larger corpus lacking such information. After training, the acquired form and meaning features can be utilized to direct the co-speech gesture algorithm. We evaluate our approach with objective measures and in a human interaction study, demonstrating its effectiveness in generating gestures that match the communicative intent and person perception of human ground truth behavior. A video with examples of generated co-speech gestures is made available online\footnote{\href{https://vimeo.com/821576001}{https://vimeo.com/821576001}}.

In summary, this work makes three main contributions. First, we introduce a novel augmented co-speech gesture synthesis framework, which can be effectively trained using a limited amount of annotated data. It generates not only natural but controllable gestures, enhancing the synthesis process. Second, we demonstrate that the framework generates high-quality co-speech gestures surpassing current state-of-the-art techniques and closely resembling those produced by humans. Finally, we present a method to analyze the impact of individual target features on the synthesized gestures, to gain a deeper understanding of the additional factors influencing co-speech gesture generation.
\section{Related Work}

\subsection{Automatic Gesture Generation}
\begin{figure}[b]
  \centering
  \includegraphics[width=\linewidth]{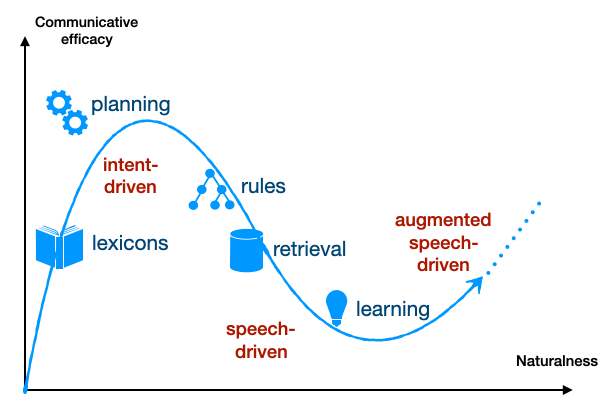}
  \caption{Different methods for co-speech gesture generation, placed according to gesture naturalness and communicative efficacy (adaptability). The blue curve illustrates the field's progression over the last three decades. We present an approach to augmented speech-driven gesture generation.}
  \label{fig:gesture_history}
\end{figure}

Traditionally, gesture generation was considered a sub task in generating a multimodal utterance for an intended communicative function. Such intent-driven approaches implemented the mapping either by selecting gesture templates from a lexicon according to hand-crafted rules or by composing them via behavior or animation planning (cf.~Fig.~\ref{fig:gesture_history}). The seminal BEAT toolkit \citet{beat_framework} used extensible rules derived from linguistic analysis of human conversational behavior. \citet{kopp_framework} proposed the Behavior Markup Language (BML) for defining mappings from functions to multimodal behavior. Other early work implemented behavior planning based on grammars \citet{tepper2004content} or probabilistic inference \citet{bergmann2009gnetic,neff2008gesture}. Later approaches applied kernel-based probabilistic models to learn patterns from motion capture sequences \cite{brand2000style} or Markov models to learn structures from temporal segmentation \cite{galata2001learning}. These techniques were employed in the first commercially available products, such as the Pepper or Nao robots, due to their high degree of customization and low computational requirements \cite{pandey2018mass}.

With the advent of deep learning and large amounts of data on human communicative behavior, learning-based methods became the predominant approach to synthesizing behavior for human-human and human-agent interaction tasks \cite{yu2019interactive,tapus2019perceiving}. This direction has predominantly targeted speech-driven gesture synthesis, i.e. the problem of synthesizing gestures for a given spoken utterance possibly along with other parameters such as speaker id, style, or contextual factors.
Multiple approaches tried to generate gestures from speech audio data and initial gesture input \cite{ferstl2019multi,ginosar2019learning,zhou_gesturemaster_2022}. \citet{ferstl2018investigating} employed a recurrent neural network with an encoder-decoder architecture trained on prosodic features, for the purpose of generating brief motion sequences. 
\citet{henter2020moglow} employed a network that utilizes invertible 1x1 convolutions to generate co-speech gestures, while \citet{ling2020character} adopted trained a variational autoencoder with deep reinforcement learning to enable goal-directed control over the gestures. 
Recent studies demonstrate that general adversarial networks (GANs) can produce highly realistic outcomes. These networks can seamlessly and cohesively integrate multiple modalities, resulting in a more comprehensive representation of human communication \cite{yoon2020speech,nyatsanga2023comprehensive,ahuja2020style}.
In an effort to merge audio, text, and speaker identity with a GAN framework, \citet{yoon2020speech} developed a technique to produce hand gestures that closely resemble those exhibited during actual human communication. This approach yields gestures that combine information from the text and the rhythm of spoken utterances.
\citet{ao2022rhythmic} recently proposed an approach to incorporate both rhythmic and semantic aspects into co-speech gesture synthesis. Specifically, they combined a segmentation pipeline based on rhythm with neural embeddings of speech and motion, informed by linguistic insights. All of these speech-driven and co-speech driven approaches manage to synthesize gestural motions that look natural with the accompanying speech, but they are largely restricted to non-representational "motor gestures" like beats.

\subsection{Gesture Adaptability and Control}
While earlier hand-crafted and statistical approaches failed to produce natural-looking gestures, they allowed for direct adaptation of the synthesized gestures to given functions, speakers, or styles \cite{kopp_framework,beat_framework,neff2008gesture}, thereby ensuring communicative efficacy (Fig.~\ref{fig:gesture_history}). Data-driven approaches, on the other hand, provide a high degree of naturalness of gestural motion, but they lack flexibility and control over the generated output \cite{ghorbani2022zeroeggs}. Consequently, increasing the adaptability of data-based approaches has moved into the focus of recent work  \cite{habibie2022motion,ghorbani2022exemplar,ghorbani2022zeroeggs,yoon2021sgtoolkit}.
\citet{ahuja2020style} devised a technique for learning style embeddings, which enabled the simulation of gesture styles for individual speakers and allowed for variation that could be dynamically adjusted. Similarly, \citet{ghorbani2022exemplar} employed a variational framework to learn a latent space and extract motion styles from motion clips, enabling the adaptation and blending of existing styles. The SGToolkit \cite{yoon2021sgtoolkit} learns fine-grained pose control and coarse style controls, allowing for the modification of speed, spatial extent, or handedness of gestures. 
\citet{habibie2022motion} combined database matching and deep generative modeling to provide direct control through search in a k-Nearest Neighbor (k-NN) space. 
Likewise, \citet{alexanderson2020style} established a framework that permits the adjustment of overall gesture level, speed, symmetry, and spatial extent of the generated gestures.  \citet{kucherenko2021speech2properties2gestures} proposed a method to first predict gesture properties and then use this to condition the generation of gestures to be more diverse and representational. However, a framework that affords both naturalness and fine-grained control over single gestures, as needed for tailoring multimodal behavior to the specific interaction context shaped by, e.g., intended communicative functions, common ground, or inter-agent alignment, is still missing. In this paper we propose to augment (learning-based) speech-driven gesture synthesis by including target form and meaning features in the generation process.

\section{Data} \label{data-acc}
\subsection{SaGA Corpus} \label{saga}
Integrating form and meaning features into data-based gesture generation requires a data corpus with certain characteristics. Firstly, it needs to contain videos or available tracking data that can be used to capture gesture sequences. Secondly, it needs a certain level of annotated labels to allow for the creation of form and meaning features. We adopted the Bielefeld Speech and Gesture Alignment Corpus (SaGA) \cite{lucking2010bielefeld}, which consists of 25 video recordings of direction-giving dyads in German. In this task, a direction-giver provides a detailed description of a city route s/he has previously seen in Virtual Reality, to another person who has to find the way afterward. The entire corpus spans 4 hours and 38 minutes, encompassing 1764 gesture sequences with comprehensive transcription and annotation of gesture-related information (gesture phases, phrases, practices, meaning, hand shapes, hand orientations, wrist locations, and movement features).

We performed high-quality 3D upper-body tracking on this corpus using the data preprocessing pipeline by \citet{quant_fram}, based on the method presented by \citet{yoon2020speech}. 59 3D full-body keypoints were tracked for each individual, using the AlphaPose library \cite{alphapose}. First, a YOLOv3 model \cite{redmon2018yolov3} is used for detecting a person in each frame of the video. Subsequently, the 2D position of each keypoint is estimated through the implementation of a FastPose model \cite{zhang2019fast}. The tracked keypoints, without the hands, are then transformed into 3D coordinates by employing the VideoPose3D model \cite{pavllo20193d}. For the hand keypoints, the pipeline adopts the MediaPipe Hands model \cite{zhang2020mediapipe} to elevate the 2D keypoints and add their corresponding 3D positions. To establish a consistent input pipeline, we translated the German transcriptions to English, with a German to English translator service, while keeping the original sentence timings intact. During the preprocessing stage, the pipeline implements various quality control measures to eliminate low-quality data. However, given the limited corpus size available for analysis, we deactivated those quality control measures, as they would have led to the exclusion of four videos.

To identify form and meaning features, we categorized the annotations in the SaGA corpus into 17 distinct groups. Whenever possible, we differentiated between features pertaining to a single hand and those involving both hands. The groups containing single-hand information are: "Gesture Phase", "Gesture Phrase", "Gesture Relative Position", "Hand Shape", "Wrist Position", "Movement Extent", and "Gesture Practice"; the groups without specific hand information are: "Spoken Entity", "Spoken Relative Position" and "Entity Occurrence".
Since the corpus only contains general categories for the meaning of a gesture, we created the "Spoken Entity", "Spoken Relative Position" and "Gesture Relative Position" labels, which provide information on whether the gesture conveys a relative position or an entity description. In addition, we added an "Occurrence" label indicating whether such a semantic feature is present or not. Specifically, the label is initially set to zero and, upon detection of an entity or a relative position label within the next two seconds, its value undergoes a linear increase to one, stays at one for the time of the detected label, followed by a linear decrease back to zero. In cases where multiple labels are present simultaneously, the higher value among the two is used.

To obtain more specific semantic gesture features, we first manually classified all nouns relevant to entities and all adjectives and adverbs relevant to position into distinct groups. We established 18 distinct groups for nouns and 13 for adverbs and adjectives. Given the presence of an entity, relative position, or gesture relative position label, we examined the one-second window before and after the instance against our designated groups and return the corresponding label. In cases where multiple labels are identified, we chose the majority vote. If a tie occurs, we refrained from returning a label to prevent the introduction of additional noise to the data. Using this method we were able to classify 96\% of the entity and position labels into our groups. We removed the remaining features from the data.

\begin{figure*}[ht]
  \centering
  \includegraphics[width=\textwidth]{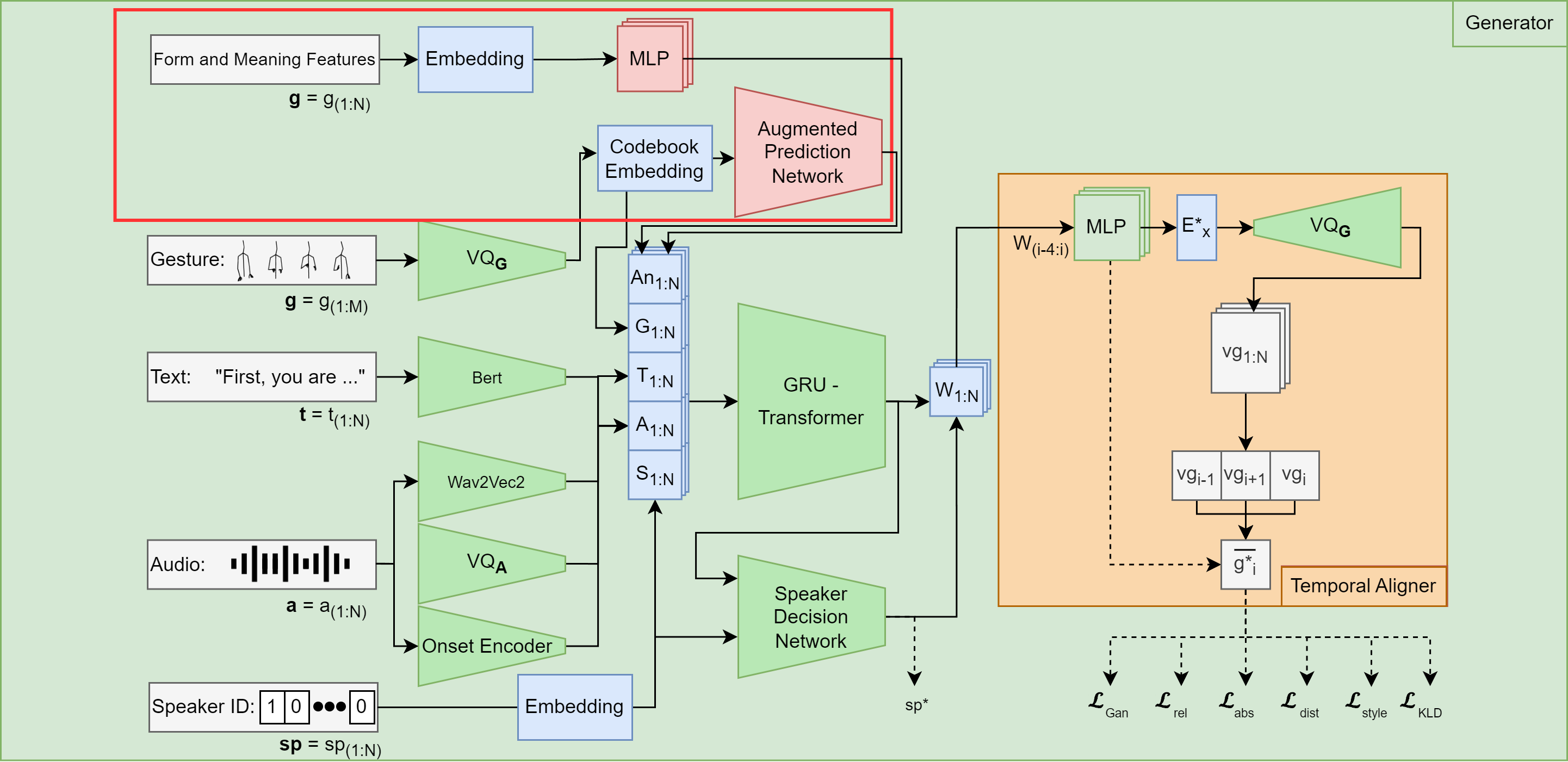}
  \caption{Overview of the framework architecture. The left-hand side shows the different input modalities, which are combined and trained in the GRU-Transformer. Temporal Aligner is marked in orange. Our extensions are marked in red.}
  \label{fig:diagram_model}
\end{figure*}

\subsection{BiGe Corpus} \label{ted_data}
Modern data-driven approaches require vast amounts of training data to learn the underlying features and patterns. 
As the SaGA corpus is relatively small, we additionally leveraged the preliminary BiGe corpus \cite{quant_fram} containing 1021 hours of videos featuring full-body 3D keypoint tracking. This corpus was collected from TED \cite{ted_youtube} and TEDx \cite{youtubeTEDxTalks} YouTube channels and provided us with sufficient data.
We relied on the pre-processed dataset provided by \citet{quant_fram}, consisting of 54,360 sequences referred to as shots of interest (SoTs), with an average length of 17 seconds. To create the SoTs, each collected video was split into multiple snippets based on camera cuts. The snippets containing low-quality and noisy tracking data were removed. After filtering, the dataset had a total duration of 260.6 hours. We divided the dataset into 208 hours of training data and 26 hours of validation- and test data to evaluate the performance of our deep learning algorithms.

\section{Augmented Gesture Synthesis}
\subsection{AQ-GT framework}
We build our augmented synthesis framework upon the work recently presented by \citet{quant_fram}. As shown in Fig.~\ref{fig:diagram_model}, it employs a hybrid gated recurrent unit (GRU) and Transformer architecture, which is trained using a GAN employing Wasserstein divergence (WGAN-div). The framework takes into account various input modalities, including prior gestures, text, speech, and speaker identity, to predict gestures frame by frame.
The prior gestures input is processed using a hierarchical Vector Quantized Variational AutoEncoder (VQ-VAE2) \cite{razavi2019generating} pre-trained with a WGAN-div approach to encode and reconstruct a sequence of up to four frames. In the framework, this model is denoted as VQ\textsubscript{G}. 
Regarding the input text, the BERT transformer model \cite{devlin2018bert} is used to construct a vector embedding for the designated text sequence.
The speech input is preprocessed in three steps. First, a custom VQVAE2 is used, which is called the VQ\textsubscript{A}. Next, the wav2vec 2.0 model \cite{baevski2020wav2vec}, primarily trained on spoken dialogue, is applied. Lastly, an Onset Encoder architecture \cite{liang2022seeg} is used to learn accurate beat segmentation. 
Speaker identity is included through an embedding lookup layer and subsequently refined via a speaker decision network with a contrastive learning strategy. 
The aggregated features are passed through a GRU-Transformer architecture and then decoded via the decoder portion of the VQ\textsubscript{G} into a short sequence before finally being refined through a temporal aligner network.

\subsection{Augmentation by Form and Meaning}
As shown in Figure \ref{fig:diagram_model}, we extend the framework by incorporating form and meaning features as an additional input modality. To that end, we utilize the form and meaning features defined in Section \ref{data-acc} as 17 categorical class labels of length $N$ and combine them with the other modalities. For this, we generate an embedding layer for every conceivable state of our categorical class labels. Throughout the training process, the categorical labels serve as reference values, allowing us to select a specific embedding for each label. After passing these embeddings through a multilayer perceptron (MLP), we employ the "Reparameterization Trick" as proposed by \citet{kingma2013auto}, to establish a multivariate normal distribution within the latent space. This technique has been shown to yield a more meaningful sampling space that later can be used to generate novel gestures \citet{ahuja2020style,ghorbani2022exemplar}. If there is no label available, such as when the framework is trained on a sample of the BiGe dataset, we set the input label to $-1$ and generate a zero-valued vector for that label.

In addition, we introduce an \textit{augmented Prediction Network} (aPN) which is pre-trained on form and meaning features. For this, we train a GRU on four input frames by passing the data through the VQ\textsubscript{G} and using the ID output as a lookup vector for an embedding layer. This embedding space is subsequently trained with a GRU to forecast the categorical class label for each of the four frames through a cross-entropy loss function. During the training of the overall framework, the aPN is frozen and used to predict form and meaning features for the first $M$ frames of the gesture input. Note that this extension is not meant to consistently generate gestures for specific labels. Rather, the idea is to address a particular challenge that hinders the generation of consistent and coherent gestures: As the framework creates gestures in chunks of $\frac{M+N}{S}$ seconds, the first frames of each generated sequence are the most difficult to create coherently, as there is limited information about the previous segment. This limitation results in the framework frequently discontinuing gestures at the onset of a newly generated segment and reverting to a resting position. Our extensions returns a probability distribution over the 16 possible label categories, which enables the framework to keep up coherence with the previously generated gesture segments.

\subsection{Training}
The augmented framework was trained on both the SaGA corpus and the BiGe corpus. We randomly split the 25 videos of the SaGA corpus into 21 videos for the training-, 2 videos for the validation- and 2 videos for the test set. For the BiGe corpus, we use the same training, validation, and test split as in \cite{quant_fram}.
We trained our framework utilizing a batch size of 190, a learning rate of $2 * 10^{-5}$, a regression weight of 20, and a GAN weight of 2. We used a sampling rate $S$ of 15 frames per second, set the number of generated poses $N$ to 30, and set the number of preceding input frames $M$ to 4. The framework was therefore trained to generate two-second data sequences. With a random chance of 25\%, we omitted the entire feature label of a sample, to force the network to also learn gestures without relying on form and meaning features. To reduce overfitting, we used dropout with a stepwise increment of 0.05 every 25 epochs, up to a maximum of 0.3, between every layer and for the input of our network.

Using the same unrestricted epoch count with an early stopping callback as in the original framework, we restored the optimally performing epoch upon completion of training. We trained the augmented framework on 187 epochs incorporating both datasets. 
For the validation and test set, we computed the loss for the BiGe and SaGA datasets separately and then averaged the result. This ensures both datasets have equal weight in the overall validation and test loss. Given that the BiGe dataset is significantly larger, we increased the frequency of the SaGA training data to ensure that 1 out of every 5 training samples originates from this dataset. This choice of over-sampling size was determined via a hyperparameter search, in which we evaluated the validation loss of the framework after 10 epochs for ratios ranging from 1:10 to 1:1. 

\begin{table}[b]
    \caption{Comparison of our system with other state-of-the-art frameworks on the BiGe dataset. For FGD and MAJE lower is better, and for Diversity higher is better.} 
  \centering
  \begin{tabular}{lccccc}
    \toprule
    Methods & FGD $\downarrow$ & Diversity $\uparrow$ & MAJE $\downarrow$\\
    \midrule
    Ground Truth & 0.00 & 42.128 & 0.00\\
    \midrule
     Trimodal \cite{yoon2020speech} & 2.023 & 38.980 & 0.0119\\
     SEEG \cite{liang2022seeg}  & 2.018  & 38.660 & 0.0121\\
     HA2G \cite{liu2022learning}   & 1.346  & 41.332 & 0.0096\\
     AQ-GT \cite{quant_fram}   & 0.3977 & 42.885 & \textbf{0.0078}\\
     \midrule 
     \textbf{Ours} & \textbf{0.3429} & \textbf{44.271} & 0.0081\\
    \bottomrule
  \end{tabular}
  \label{tbl:res}
\end{table}

\section{Evaluation}
We evaluated the proposed augmented gesture synthesis framework using three quantitative metrics: the Fréchet Gesture Distance (FGD), the mean absolute error between generated joint positions (MAJE), and the average feature distance (Diversity). Please note that the MAJE is mainly reported for legacy sake, as this metric is unable to account for the one-to-many mapping problem in gesture generation.
FGD \cite{yoon2020speech} measures the similarity between latent vectors of a generated gesture and a ground truth gesture, by computing the ordered point distance between them. We obtain the latent vectors from the higher-dimensional Embedding Network version of the original framework \cite{quant_fram}.
The average feature distance \citet{lee2019dancing} scores the diversity of a set of generated gestures, by measuring the absolute distance between random pairs of gesture samples based on their latent representations. Due to the random pairing, the distance may vary significantly between each measurement. Therefore, we estimate the Diversity distance as the average of one thousand calculations.

Table \ref{tbl:res} shows a comparison of the results for our framework with other state-of-the-art frameworks, including the base AQ-GT framework. As the other frameworks were trained on the BiGe corpus, we report the results based only on the test split of the BiGe corpus. The results indicate improvements in both FGD and Diversity Score. That is, augmenting the framework to take into account form and meaning features improves the framework's ability to produce more human-like and more diverse co-speech gesturing also on the BiGe test set.
Given the lower FGD, the higher Diversity score and the deterioration of the MAJE is to be expected, as adding the SaGA data would naturally lead the framework to synthesize gesture samples that are outside of the BiGe dataset distribution. This indicates, nevertheless, that the framework has learned to incorporate the SaGA dataset and is able to generate more diverse gestures.

Analyzing the results from the SaGA test split, we find that FGD is 1.336, Diversity score is 39.198 and MAJE is 0.0121. Compared to the BiGe data, these results are inferior in all respects, which was expected as the SaGA dataset contains more complex, representational gestures while being much smaller in size. While other splits between both corpora produced results more favorable for the SaGA test, overall framework performance was reduced due to the high level of noise in the SaGA dataset.

\subsection{Guidance by Target Features}
\begin{figure}[tb]
  \centering
  \includegraphics[width=\linewidth]{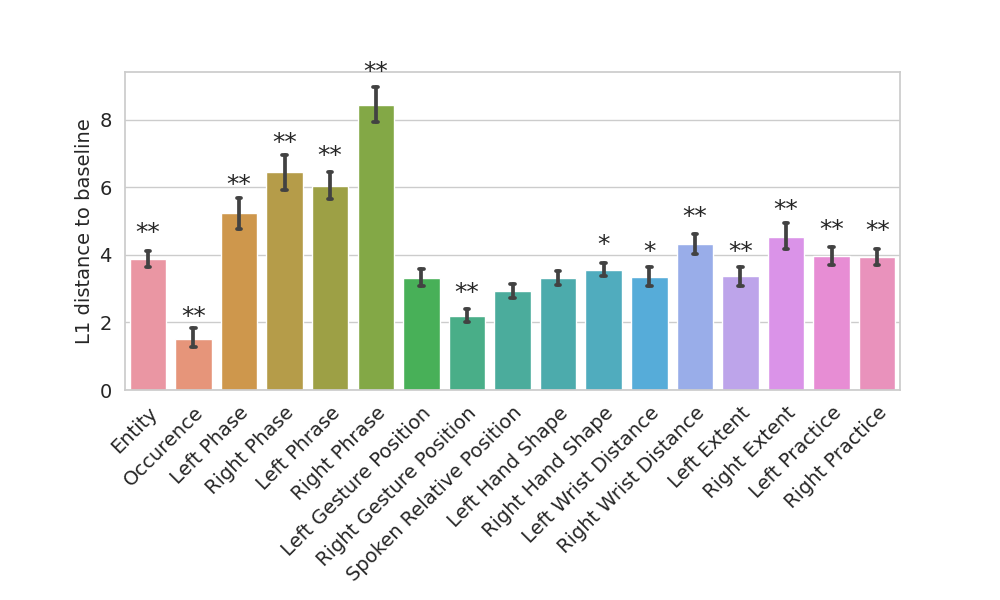}
  \caption{Mean L1 distance between gestures synthesized with the original features (baseline) and individually altered form and meaning features, presented with 95\% CI. Asterisks indicate significant effects (\text{*} : p < 0.05, \text{*}\text{*} : p < 0.005).} 
  \label{fig:variance}
\end{figure}

We investigated how gesture generation can be guided by single form and meaning target features, by conducting a variance analysis for each discrete feature label. As a baseline, gesture sequences were generated for the entire test set at 20-second intervals with the original form and meaning labels. 
Afterward, the co-speech gestures were synthesized for the same intervals but with every categorical feature label systematically varied. To ensure consistent generation for each sequence, we set all available random seeds to zero and deterministic, if applicable. To compare the synthesized gestures with the baseline gestures, the mean and variance of the L1 distance were calculated (see Fig.~\ref{fig:variance}).
    
Since the data does not meet the normality and homogeneity of variance assumptions, we applied a Generalized linear mixed-effects model (GLME) with a Gaussian distribution to test the differences. Further, given that the data is continuous and strictly positive, we also chose to use a log link function to stabilize the data.

The results indicate that modifying single features as input indeed resulted in distinct variations of the gesture output. Altering the gesture's "Phase" and "Phrase" led to the most significant differences, while changing "left relative gesture position", "spoken relative position" and "left-hand shape" did not produce a significant difference in the L1 distance. Surprisingly, two meaning features, "Left Gesture Position" and "Spoken Relative Position" did not have a significant impact on the generated gestures, despite showing differences when inspecting the generated videos. 
Notably, for gesture position, hand shape, and wrist distance, the differences are more significant for the right hand only. This could be due to the fact that right-handed gestures are more frequent in the SaGA training data (ratio 1 : 1.39) or that there are more expressive right-handed gestures in the SaGA corpus.

\subsection{Human Rater Study}

\begin{figure*}[tb]
  \centering
  \includegraphics[width=17cm]{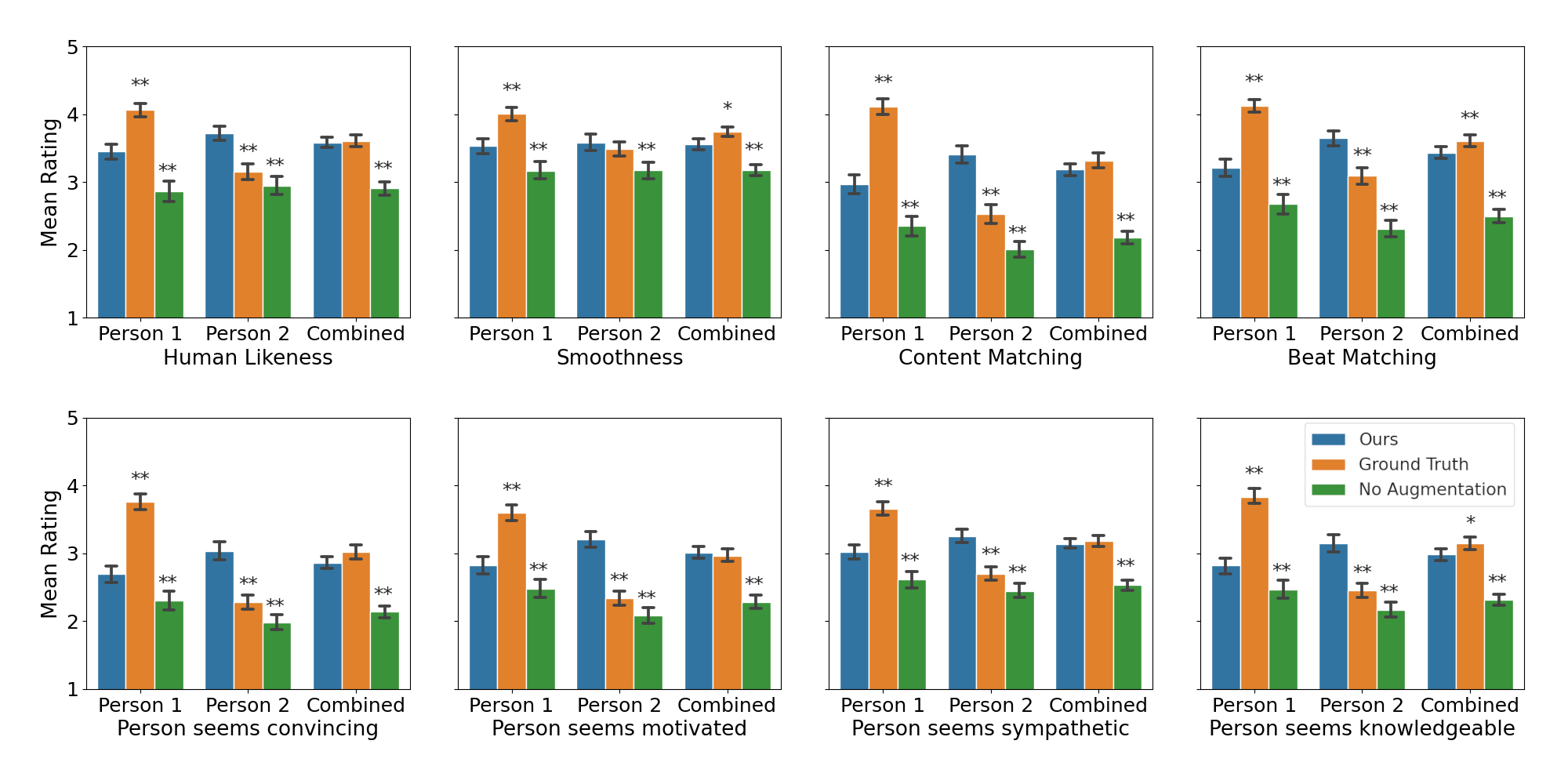}
  \caption{Results of the human rater study. The augmented framework is shown in blue, the human ground truth is in orange, and the AQ-GT framework is in green. Asterisks indicate significant effects (\text{*} : p < 0.05, \text{*}\text{*} : p < 0.005).}
  \label{fig:study_result}
\end{figure*}

A second common evaluation method for gesture generation models is to have human raters subjectively assess the quality of the synthesized output.
To that end, we generated 20-second sequences for the entire SaGA test set, each consisting of the results synthesized by our augmented network, the ground truth, and the AQ-GT framework \cite{quant_fram}. For a fair comparison, we used the re-trained weights for the AQ-GT framework, but without any form and meaning labels and without the aPN extension. We randomly selected 10 videos for an online evaluation study with 70 German-speaking participants (50\% female and 50\% male). Participants had to watch the videos in random order and, using a 5-point Likert scale, rate the co-speech gestures concerning human likeness and smoothness of motion as well as how much it matched the accompanied speech content and timing (beat). In addition, we asked participants how they perceive the gesturing person by having them enter how much they agree/disagree with the following statements (translated into English): "The gesturing makes the person seem convincing", "The gesturing makes the person seem motivated", "The gesturing makes the person seem sympathetic" and "The gesturing shows that the person has understood what he or she is expressing".
As each video was 20 seconds long, we excluded any participants who remained less than 20 seconds on the survey pages, as well as any participants that deviated more than two times the standard deviation from the mean. After applying these conditions, 60 participants remained. 

We used a generalized linear mixed-effects model (GLMM) for a repeated measure analysis. To compare the pairwise significance effect of the different frameworks, we then used the Dunn test, which does not assume the normality of the data or equal variances across groups. 
Figure \ref{fig:study_result} shows the ratings of co-speech gestures from human ground truth (orange), synthesized with the augmented framework (blue) or the re-trained AQ-GT framework (green), for input from two specific speakers in the SaGA test set.
These speakers were selected randomly, ending up with "Person 1" who frequently employed elaborate, rapid, and iconic hand gestures, while "Person 2" primarily and infrequently utilized small beat gestures.

Comparing the augmented framework to the re-trained AQ-GT framework, we see significantly improved ratings in all statements and for each speaker input. The large increase in the overall rating of the augmented framework clearly shows that our framework improved the gesture synthesis over the previous framework and learned to incorporate the form and meaning features from the SaGA corpus. In addition, the gesture sequences generated by the augmented framework are rated comparable to human ground truth with respect to human likeness and content matching. This suggests that the augmented framework has successfully incorporated form and meaning features and generates gestures that, on average, match the spoken context as well as those produced by humans. However, for the statements "Smoothness" and "Beat Matching," the augmented framework's gestures are rated significantly lower, indicating that it may generate some jerky movements to comply with the given features. 

Considering the results for the individual speakers, we observe that the augmented framework receives significantly lower ratings than the human ground truth of "Person 1". However, except for "Smoothness," it gets significantly higher ratings than the ground truth of "Person 2". This may indicate that the framework has learned to incorporate representational (iconic and deictic) gestures into the normal, mainly beat-focused gesture synthesis. Yet, the framework is unable to produce complex gestures that are as believable as those produced by "Person 1". 
Similarly, when looking at the effects on person perception (bottom row in Fig.~\ref{fig:study_result}), the output generated by our framework is rated significantly lower than "Person 1" but significantly higher than "Person 2". In terms of the combined results this leads to the augmented framework receiving ratings similar to human ground truth, except for the statement "Person seems knowledgeable". That is, there are clear differences in the perception of gesturing individuals and the synthesized gestures can well be on par with those produced by humans, but still may not achieve the full expressiveness of human communication. 
Further, although our framework can incorporate features and generate complex gesture sequences, it may struggle to adjust to sudden quick rhythm changes. This could explain the higher average ratings for "Person 2", who uses small and more monotonous gestures, allowing our framework to better adjust to the speaker's rhythm and generate more convincing gestures. 

A caveat is in order here, though. In our study design, participants rated different gesture sequences with the same audio, which could have led to an anchoring bias \citep{FURNHAM201135}. That is, participants may have based their ratings on the first gesture sequence they saw and adjusted subsequent ratings based on this initial rating. Although we randomized the order of the sequences, it is possible that this was not enough to fully exclude carry-over effects. 
Comparing these study results with the results of the original framework, we can observe a lower average score for the AQ-GT framework, which can be explained by the much higher gesture expressivity of the person in the SaGA dataset, compared to the person in the BiGe dataset.

\begin{table}[b]
  \centering
  \caption{Comparison of our augmented network w/ and w/o the form and meaning features and the aPN. For FGD the lower the better, and the higher the better for the Diversity score.}
  \begin{tabular}{lcccc}
    \toprule
    Methods & FGD $\downarrow$ & Diversity $\uparrow$ & MAJE $\downarrow$\\\
    \textbf{Ground Truth} & 0.00 & 41.690 & 0.00\\
    \midrule
    \textbf{w/o Features \& aPN} & 2.207 & \textbf{39.747} & 0.0139 \\
    \textbf{w/o Features} & 1.791 & 39.538 & 0.0136\\
    \textbf{w/o aPN} & 1.546 & 39.128 & 0.0130\\

    \midrule
    \textbf{Original} & \textbf{1.336} & 39.198 & \textbf{0.0121}\\ 
    \bottomrule
  \end{tabular}
  \label{tbl:ablation}
\end{table}

\section{Ablation study}
An ablation study was performed to assess the effect of the added components on the performance of the augmented network. Specifically, we evaluated the impact of removing the feature input, setting the weights of the aPN to zero, or both, on the overall performance of the network. We re-train each architecture for 50 epochs and report the FGD, Diversity score, and MAJE of the SaGA test split in table \ref{tbl:ablation}. As can be seen, removing any of the added components has a clear adverse effect on the FGD value and the MAJE of the augmented framework. In contrast, removing any of the components increases the Diversity Score, and removing both yields the highest Diversity score. Although this is surprising, it can be explained by our previous observations. 
As the diversity score only takes the generated latent vectors inside a distribution into account and not the performed gestures, erratic or quick gestures can increase the diversity score. 
For our framework, we specifically designed the aPN to remove the quick recession to a resting position during the onset of a new gesture sequence and enable the continuation of the previous gestures. This design choice was made to ensure smooth and continuous gesture sequences. The removal of this component likely resulted in more erratic gestures and thus a higher diversity score. The increase with the removal of the form and meaning features can be similarly explained. As these features are designed to guide the augmented network to perform certain gestures at specific points in time, the removal of those features could lead to more frequent switching between gestures and overall more erratic gesturing. If the removal of the features and the removal of the aPN are combined, this effect can be even more exaggerated.

\section{Conclusion}
We have presented a novel co-speech gesture synthesis framework that aims to increase the range and appropriateness of generated gestures by incorporating target form and meaning features. We showed that our framework acquires the corresponding information during training on the densely annotated SaGA corpus, together with the larger BiGe corpus that does not include any annotations. A variance analysis indicates a clear influence of single features (given as additional input) on the gesture synthesis process, showing differences between the generated gestures with a significant impact on handedness. Even more interestingly, the subjective evaluation study showed that on average the augmented framework is on par with human ground truth gestures, both in terms of naturalness and how the gesturing person is perceived. Despite these positive results, however, there remains much room for improvement. According to the individual results of the subjective evaluation, our framework still struggles to generate distinctively representational iconic and deictic gestures from the features provided. Likewise, it has difficulties adapting it motion dynamics to abrupt changes in speech rhythm.  
Our study only scratched the surface of what can be done with this approach and we plan to follow up by investigating the influence of higher-level features on the training and inference properties of our framework. As a first step, we aim to conduct a more comprehensive assessment of each distinct feature label. This exploration will allow us to investigate how these features can be manipulated to create novel co-speech gestures that can be utilized in a variety of real-world settings. Additionally, we plan to examine whether features from the SaGA dataset can be leveraged to enhance other English datasets, ultimately improving the accuracy and controllability of generated co-speech gestures.

\clearpage
\bibliographystyle{ACM-Reference-Format}
\bibliography{main}


\end{document}